\documentclass[12pt]{article}
\usepackage{amsmath, amssymb}
\usepackage[dvips]{graphicx}

\begin{document}
\baselineskip=16pt

\begin{titlepage}
\begin{flushright}
{\small IPMU-11-0011}\\
{\small OU-HET 691/2010}\\
{\small UT-11-04}\\
\end{flushright}

\vspace{1.5cm}

\begin{center}

{\Large\bf Sneutrino Inflation with Asymmetric Dark Matter} \\
\vspace{0.3cm}

\vspace{2.5cm}

{\large Naoyuki Haba}$^1$,
{\large Shigeki Matsumoto}$^2$,
and
{\large Ryosuke Sato}$^{2,3}$

\vspace{1cm}

$^1${\it Department of Physics, Osaka University, Toyonaka, 560-0043, Japan}\\
$^2${\it IPMU, TODIAS, The University of Tokyo, Kashiwa, 277-8583, Japan}\\
$^3${\it Department of Physics, The University of Tokyo, Tokyo, 113-0033, Japan}\\

\vspace{3.0cm}

{\bf Abstract}\\
\vspace{0.5cm}
{\parbox{13cm}{\hspace{5mm}
The asymmetric dark matter scenario is known to give an interesting solution for the cosmic coincidence problem between baryon and dark matter densities. In the scenario, the dark matter asymmetry, which is translated to the dark matter density in the present universe, is transferred from the $B-L$ asymmetry generated in the early universe. On the other hand, the generation of the $B-L$ asymmetry is simply assumed, though many mechanisms for the generation are expected to be consistent with the scenario. We show that the generation of the asymmetry in the sneutrino inflation scenario works similarly to the asymmetric dark matter scenario, because the nonrenormalizable operator which translates the $B-L$ asymmetry to the dark matter asymmetry is naturally obtained by integrating right-handed neutrinos out. As a result, important issues concerning cosmology (inflation, the mass density of dark matter, and the baryon asymmetry of the universe) as well as neutrino masses and mixing have a unified origin, namely, the right-handed neutrinos.}}

\end{center}

\end{titlepage}

\section{Introduction}

Thanks to recent cosmological observations such as the WMAP (Wilkinson Microwave Anisotropy Probe) experiment~\cite{Komatsu:2010fb}, we now know the energy budget of our Universe; the baryon density of the universe is $\Omega_B \sim 0.04$, while that of dark matter (DM) also is in the same order, which is five times larger than $\Omega_B$ ($\Omega_{DM}\sim 5\Omega_B$). One of the interesting explanations of its origin is found in the ``asymmetric dark matter (ADM)'' scenario~\cite{Kaplan:2009ag}, where the existences of both dark and antidark matters are postulated. Dark and antidark matters are interacting with standard model (SM) particles in the early universe, which leads to $DM \sim B$ with $DM$ and $B$ being asymmetries of dark matter and baryon, so that above relation between $\Omega_{DM}$ and $\Omega_B$ holds naturally when the mass of the dark matter is in an appropriate order. The ADM scenario can be realized when there is a messenger interaction which transfers the baryon (lepton) number to the dark matter number. Because of this excellent explanation between dark matter and baryon densities, several studies have been performed so far~\cite{Cohen:2010kn}-\cite{Falkowski:2011xh}. A similar scenario has also been proposed in a context of technicolor-like setup~\cite{Kribs:2009fy}.

In the ADM scenario, the $B - L$ (baryon $-$ lepton) asymmetry is simply assumed to be generated by some mechanism, which is distributed to baryon and dark matter asymmetries through sphaleron and ADM messenger interactions. Several mechanisms to generate the asymmetry such as the leptogenesis~\cite{Fukugita:1986hr} are expected to be suitable for the ADM scenario~\cite{ Falkowski:2011xh}. Among those, we show that the $B-L$ generation in the sneutrino inflation scenario~\cite{Murayama:1992ua} works similarly to the ADM scenario. The sneutrino inflation is known to be one of the most attractive scenarios to generate the $B-L$ asymmetry, because it allows us to explain not only the asymmetry but also inflation and neutrino masses and mixing, simultaneously. Interestingly, it can be shown that, when we simply add the dark and antidark matter fields to the Lagrangian of the sneutrino inflation, the most important ingredient of the ADM scenario, namely the ADM messenger interaction can be derived naturally as a nonrenormalizable operator suppressed by an appropriate energy scale. As a result, important cosmological issues (inflation, dark matter and baryon mass densities of the universe) and neutrino masses and mixing have the same origin in the model of sneutrino inflation with the ADM.

In what follows, we first show our setup, where the Lagrangian of the sneutrino inflation with the ADM fields is given. In the construction of the Lagrangian, the $Z_{4R}$ discrete symmetry, which is the subgroup of $U(1)_R$, is imposed in order for the ADM messenger interaction to play a crucial role. We next consider the thermal history of our Universe described by the Lagrangian and show that the dark matter and baryon asymmetries in the present Universe are generated in this framework. Though the purpose of this paper is to give an idea of the sneutrino inflation with the ADM, we also present some numerical results to show that the scenario does work quantitatively.

\section{Setup}

We first show the setup of our scenario, where the SM gauge singlet fields $X$ and $\bar{X}$ are introduced. Fermionic components of these superfields are nothing but dark and antidark matter fields, respectively. On the other hand, we assume that their scalar components have positive soft-SUSY breaking masses and are heavy enough compared to the fermionic components. The charges of the $Z_{4R}$ discrete symmetry, which is a part of $U(1)_R$, and the lepton number of these fields are assigned as follows.
\begin{center}
\begin{tabular}{|c|cc|}
\hline
& $X$ & $\bar{X}$ \rule{0cm}{2.5ex} \\ 
\hline
$Z_{4R}$ &   $i$ &   $-i$ \\
$U(1)_L$ & $-1/2$ & $1/2$ \\
\hline
\end{tabular}
\end{center}
With the use of the charge assignments and considering a renormalizable theory (it can be, therefore, regarded as a UV completion theory\footnote{A minimum model which cannot be a UV completion theory is also considerable, whose superpotential is $\mathcal{W}= \mathcal{W}_{\rm NMSSM} - (1/2) M_{ij} N_i N_j - m X\bar{X} + h_{ij} H_u N_i^c L_j + (\kappa_i/M_m) \bar{X}^2 L_i H_u + \lambda_X SX\bar{X}$. Here, $M_m$ is not related to Majorana masses of right-handed neutrinos, and it can take any energy scale.}), the superpotential is given by 
\begin{eqnarray} 
\mathcal{W}
=
\mathcal{W}_{\rm NMSSM}
- {1\over2} M_{ij} N_i^c N_j^c 
- m X\bar{X} 
+ h^{(\nu)}_{ij} H_u N_i^c L_j
+ \frac{\kappa_i}{2} N_i^c \bar{X}^2
+ \lambda_X S X \bar{X},
\label{W}
\end{eqnarray}
where $L_i$ is the lepton doublet of the $i$-th generation ($i = 1, 2, 3$), $H_{u(d)}$ is the Higgs doublet giving the masses of up (down) type quarks, $S$ is the singlet field predicted in the next-to-minimal supersymmetric standard model, and $N_i^c$ is the right-handed neutrino with the Majorana mass ($M_{ij}$) which breaks $U(1)_L$. The $U(1)_L$ symmetry is postulated to be softly broken, so that it is broken only by $M_{ij}$. We take the basis where $m$ and $M_{ij}$ is real (orthogonal) by the redefinitions of $X$, $\bar{X}$, and $N_i^c$ fields. On the other hand, the Yukawa coupling $h^{(\nu)}_{ij}$ and $\kappa_i$ is still complex on this basis, which plays a crucial role for the generation of the $B-L$ asymmetry. The superpotential of the next-to-minimal supersymmetric standard model (NMSSM) is denoted by $\mathcal{W}_{\rm NMSSM}$.

In this setup, one of the right-handed sneutrinos is regarded as the inflaton to drive the chaotic inflation of the very early Universe~\cite{Linde}
\footnote{
In supergravity, the chaotic inflation using F-term with minimal K$\ddot{{\rm a}}$hler potential is difficult,
because the scalar potential has exponential factor.
It is known that one can get the flat potential by using non-minimal K$\ddot{{\rm a}}$hler potential~\cite{Murayama:1993xu} or
D-term inflation~\cite{Kadota:2005mt}.
In either case, the right-handed sneutrino can have large VEV, and act as the inflaton.
}.
With the Planck scale $M_{\rm pl}$ being $10^{19}$ GeV, the anisotropy of the cosmic microwave background (CMB) induced from the inflation suggests $\delta T/T \simeq 10 (M_{ij}/M_{\rm pl}) \simeq 10^{-5}$~\cite{Salopek:1992zg}. The scale of the Majorana mass is therefore fixed to be of the order of $10^{13}$ GeV. In the following discussions, we assume that all right-handed neutrinos have a common Majorana mass, $M = 10^{13}$ GeV, for simplicity. Among three right-handed neutrinos ($\tilde{N}_{i = 1, 2, 3}$), that with the smallest couplings is expected to act as the inflaton. We also assume that the right-handed sneutrino of the first generation $\tilde{N}_1$ has such a role. It is worth noting that the largest coupling of $\tilde{N}_1$ among $h^{(\nu)}_{1j}$ and $\kappa_1$ sets the initial condition of the chaotic inflation, so that these couplings should be suppressed sufficiently.

After integrating $N^c_i$ fields out from the potential in Eq.(\ref{W}), we obtain higher-dimensional operators in the superpotential describing physics below the scale $M$,
\begin{eqnarray}
\mathcal{W}_{\rm HD}
=
{y_i^2 \over 2M}(L_i H_u)(L_i H_u)
+ {\kappa_i^2 \over 8M}\bar{X}^4
+ {y_i \kappa_i \over 2M}L_i H_u \bar{X}^2,
\label{HD}
\end{eqnarray}
where the coupling constant $y_i$ is obtained by diagonalizing the Yukawa coupling $h_{ij}^{(\nu)}$, which is given by the neutrino mass $m_{\nu,i}$ as $|y_i|^2 \equiv 2m_{\nu,i}M/v^2 \sin^2 \beta$ with the vacuum expectation value of the Higgs field to be $v= 246$ GeV. The higher-dimensional operators change the lepton number $L$ ($\Delta L = 2$), which originates in the Majorana mass of right-handed neutrinos. Among those, the first one, $(L_i H_u)(L_j H_u)$, is the operator to create neutrino masses. Tiny neutrino masses observed by the oscillation experiments~\cite{Strumia} suggests that the mass of the right-handed neutrino ($M$) should be around $10^{13}$ GeV with $|y_i|$ being ${\cal O}(0.1$-$1)$, which is consistent with the scale required from the observation of the CMB anisotropy. The second one, $\bar{X}^4$, gives the self-interaction between antidark matters, which also changes the dark matter number and plays an important role to reduce the dark matter asymmetry to a suitable value required by observations. The last one, $L_i H_u\bar{X}^2 $, is nothing but the operator which plays a crucial role for the mediation of the $B-L$ asymmetry to the dark matter asymmetry in the early Universe. The nonrenormalizable operator of the ADM messenger interaction should be suppressed by the energy scale smaller than 10$^{14}$ GeV\footnote{The upper bound of the scale $M$ comes from the fact that the ADM interaction should not be decoupled before the Sphaleron process be activated, namely, the ADM interaction should be active, at least, at the temperature around $10^{12}$ GeV. On the other hand, if the condition $M > 10^9$ GeV holds, the ADM interaction decouples before the Sphaleron process becomes inefficient, namely, the ADM interaction is not active when the temperature of the universe is below 100 GeV.} when the coefficient of the operator is ${\cal O}(1)$~\cite{Haba:2010bm}. Interestingly, the scale of $M$ required from both the observations of the CMB anisotropy and the tiny neutrino masses satisfies the condition. As a result, the ADM messenger interaction, $L_i H_u\bar{X}^2 $, works well to transfer the $B-L$ to the dark matter asymmetries.

\section{Sneutrino inflation scenario with ADM}

We are now in a the position to discuss the thermal history of the Universe described by the model presented in the previous section. We will discuss how the inflation proceeds: the decay of the inflaton ($\tilde{N}_1$) generates entropy with the $B-L$ asymmetry, the asymmetry is distributed to the SM and dark sectors through the sphaleron and ADM messenger interactions, and the baryon asymmetry and dark matter density at the present universe are determined, based on simple analytical descriptions on each phenomenon. There are four important phenomena in thermal history: the inflation, the reheating of the universe, generations of $B-L$ and dark matter asymmetries, and determinations of dark matter and baryon densities in the present Universe.

\subsection{Inflation}

We have assumed that the right-handed sneutrino of the first generation $\tilde{N}_1$ is regarded as the inflaton in our setup. The chaotic condition for the inflation~\cite{Linde} requires that coupling constants associated with the right-handed sneutrino, $h^{(\nu)}_{1j}$ and $\kappa_1$, should be suppressed sufficiently in order for the $\tilde{N}_1$ to act as the inflaton. Since the slepton and Higgs fields die away due to the fact that they obtain masses from the expectation value $\langle \tilde{N}_1 \rangle$ which is larger than the Hubble parameter $H$, the equation of the motion for the inflaton field turns out to simply be 
\begin{eqnarray}
{\ddot {\tilde N}_1}
+3 H {\dot {\tilde N}_1}
+\Gamma {\dot {\tilde N}_1}
+M^2 \tilde{N}_1
= 0,
\end{eqnarray}
where $\tilde{N}_1$ is the amplitude of the inflaton field. The decay width of $\tilde{N}_1$ is denoted by $\Gamma$ and its explicit form is given by $\Gamma = M(4\sum_j|h_{1j}|^2 + |\kappa_1|^2)/(16\pi)$. The Hubble parameter is determined by the Friedman equation, $H^2 = 8\pi\rho_{\tilde{N}_1}/(3M_{\rm pl}^2)$, where $\rho_{\tilde{N}_1}$ is the energy density of the inflaton $\tilde{N}_1$. Needless to say, the inflaton is rolling down the potential until the time $H \sim M$, and it gives a sufficient e-folding number to solve flatness and horizon problems of the standard big bang cosmology.

\subsection{Reheating}

\begin{figure}[t]
\begin{center}
\includegraphics[scale=0.75]{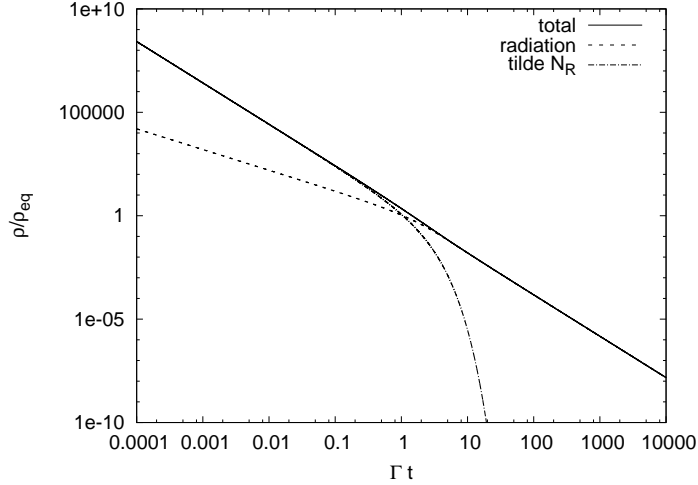}
\caption{\small Energy densities of the inflaton ($\rho_{\tilde{N}_1}$) and the radiation ($\rho_R$), and the total energy density ($\rho_{\tilde{N}_1}$ + $\rho_R$) as a function of $\Gamma t$. Here, $\rho_{\rm eq}$ is the density at the time when $\rho_{\tilde{N}_1} = \rho_R$.}
\label{rho}
\end{center}
\end{figure}

The inflaton starts to oscillate coherently around the minimum of the potential after the time $H \sim M$. The amplitude of the oscillation is gradually decreasing due to the expansion of the Universe, and eventually the inflaton decays when $H \sim \Gamma$. Decaying products make up the thermal medium through their self-interactions and the universe is reheated up to the temperature $T_{\rm RH} \simeq 0.1 (\Gamma M_{\rm pl})^{1/2}$. It is well known that the high reheating temperature often receives stringent constraints due to the gravitino problem. The relatively heavy gravitino are constrained from the big bang nucleosynthesis~\cite{Kohri:2005wn}, and the stable gravitino are constrained from the relic abundance~\cite{Moroi:1993mb}. We therefore simply assume that the gravitino mass is less than ${\cal O}(10)$eV scale to avoid the constraints. The reheating process is quantitatively described by the following equations:
\begin{eqnarray}
&&
\dot{\rho}_{\tilde{N}_1}
+3H\rho_{\tilde{N}_1}
+\Gamma \rho_{\tilde{N}_1}
=0,
\\
&&
\dot{\rho}_R~
+4H\rho_R~
-\Gamma \rho_{\tilde{N}_1}~
=0,
\end{eqnarray}
where $\rho_R$ is the energy density of the radiation originally from the decay of the inflaton. The Hubble parameter is then determined by $H^2 = 8\pi(\rho_{\tilde{N}_1} + \rho_R)/(3M_{\rm pl}^2)$. In Fig.\ref{rho}, the energy densities $\rho_{\tilde{N}_1}$ and $\rho_R$ are shown as a function of $\Gamma t$, where $\Gamma$ is set to be $10^{-12} M$. We set the time $t = 0$ when $H = M$. It can be seen that the energy density of the inflaton $\rho_{\tilde{N}_1}$ is translated to that of radiation $\rho_R$ when $t \simeq 1/\Gamma$. The total energy density $\rho_{\tilde{N}_1} + \rho_R$ is also shown in the figure, and its behavior is proportional to $t^{-2}$ (except the period $t \simeq 1/\Gamma$) due to the expansion of the Universe.

\subsection{$B-L$ and dark matter asymmetries}

The $B-L$ and dark matter asymmetries ($a_{BL}$ and $a_{DM}$) are also generated at the same time of the entropy production (the production of the radiation)\footnote{Since the Sphaleron process is active after the inflation era, the lepton asymmetry generated by the inflaton decay is immediately transferred to the $B-L$ asymmetry through the process.}, where the asymmetries are quantitatively evaluated by following two equations,
\begin{eqnarray}
\dot{a}_{\small BL}+3Ha_{BL}
=
\epsilon_{BL} \Gamma \frac{\rho_{\tilde{N}_1}}{M}
-(12 |y_i^2|^2 + |y_i \kappa_i|^2) 
\frac{231 \zeta_3 T^3 a_{BL}}{5056 \pi^3 M^2}
-|y_i \kappa_i|^2
\frac{21 \zeta_3 T^3 a_{DM}}{64 \pi^3 M^2},\label{eq:beq1}
\\
\dot{a}_{DM}+3Ha_{DM}
=
\epsilon_{DM} \Gamma \frac{\rho_{\tilde{N}_1}}{M}
-(2|y_i \kappa_i|^2 + 3|\kappa_i^2|^2) 
\frac{21 \zeta_3 T^3 a_{DM}}{64 \pi^3 M^2}
-|y_i \kappa_i|^2
\frac{231 \zeta_3 T^3 a_{BL}}{2528 \pi^3 M^2},\label{eq:beq2}
\end{eqnarray}
where $\zeta_3 = \zeta(3)$ and the temperature of the universe $T$ is given by $T^4 = \rho_R/(\pi^2 g_*/30)$ with the massless degrees of freedom $g_* = 232.5$. Parameters $\epsilon_{BL}$ and $\epsilon_{DM}$ are asymmetries generated by the decay of the inflaton $\tilde{N}_1$, which are defined as
\begin{eqnarray}
\epsilon_{BL}
=
\frac{\ln 2}{8\pi}
\frac{{\rm Im}
\left[
h^{(\nu)}_{1j}h^{(\nu)}_{1i}h^{(\nu)*}_{kj}h^{(\nu)*}_{ki}
\right]}
{|\kappa_1|^2/4 + |h^{(\nu)}_{1j}|^2},
\qquad
\epsilon_{DM}
=
\frac{\ln 2}{16\pi}
\frac{{\rm Im}\left[\kappa_1^2 \kappa_i^{*2}\right]}
{|\kappa_1|^2/4 + |h^{(\nu)}_{1j}|^2}.
\label{ASM}
\end{eqnarray}

\begin{figure}[t]
\begin{center}
\includegraphics[scale=0.8]{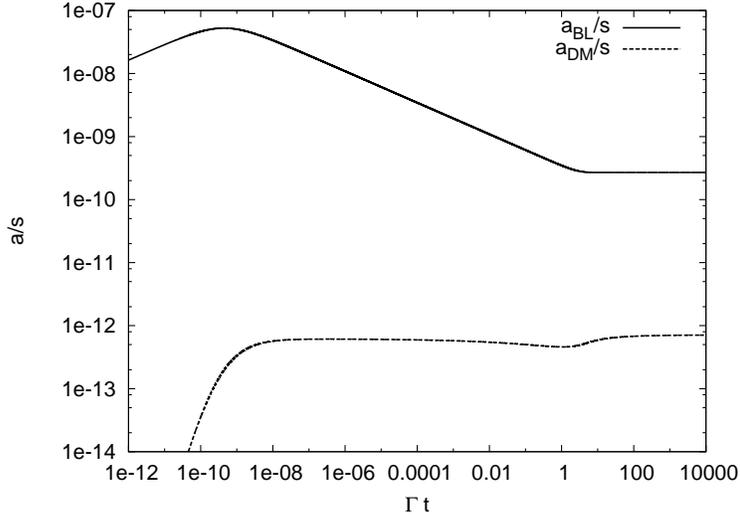}
\caption{\small Asymmetries of the entropy density ratio ($a_{BL}/s$ and $a_{DM}/s$) as a function of $\Gamma t$. The ratio $a_{BL}/s$ is shown as a solid line, while a dashed line is used for $a_{DM}/s$. For the plot, parameters $\Gamma = 3.1\times 10^3$~GeV, $\epsilon_{BL}=1.0\times 10^{-7}$, $\kappa_3 = 0.70$ and $y_3 = 0.14$ are used.}
\label{asymmetries}
\end{center}
\end{figure}

In Fig.\ref{asymmetries}, the time-development of the ratios between the asymmetries and the entropy density ($a_{BL}/s$ and $a_{DM}/s$) are shown with assuming that all asymmetries are generated by the inflaton decay into leptons, namely, $\epsilon_{BL} \gg \epsilon_{DM}$. Here, the entropy density is defined by $s = (2\pi^2/45)g_*T^3$. In this plot, parameters $\Gamma$, $\epsilon_{BL}$, and $\kappa_3$ are set to be $3.1 \times 10^3$~GeV, $1.0\times 10^{-7}$, and 0.70, respectively. Parameters $\kappa_1$ and $\kappa_2$ are assumed to be small enough compared to $\kappa_3$. We also assume that active neutrinos have the hierarchical mass spectrum in order to fix the parameter $y_i$. In this calculation, we set $y_3=0.14$ and $y_1,~y_2 \ll y_3$. It can be seen that a suitable number of dark matter asymmetry is generated by the $B - L$ asymmetry through the ADM messenger interaction. It is also worth noting that higher-dimensional operators in Eq.(\ref{HD}) give not only the messenger interaction but also washout effects on generated asymmetries as can be seen in the equations describing the asymmetries. Thanks to the effects, the generated $B-L$ asymmetry $a_{BL}/s$ can be $\sim 10^{-10}$ even if $\epsilon_{BL}$ is not so small.

\subsection{Baryon and dark matter abundances}

Once the dark matter asymmetry is generated, it is translated to the dark matter density in the present universe through annihilations between dark and antidark matter particles. When the cross section of the annihilation is larger than ${\cal O}(1-10)$ pb, either $X$ or $\bar{X}$, depending on the sign of the asymmetry $a_{DM}$, dies away from the universe~\cite{Graesser:2011wi}. In our setup, there are several annihilation processes, $X \bar{X} \rightarrow H_u H_d$ (s-channel exchange of $S$), $X \bar{X} \rightarrow SS$ (t- \& u-channel exchanges of $X$), $X \bar{X} \rightarrow \tilde{H}_u \tilde{H}_d$ (t-channel exchange of $\tilde{X}$), etc. For instance, the annihilation cross section of $X \bar{X} \rightarrow \tilde{H}_u \tilde{H}_d$ is estimated to be $\langle \sigma v \rangle \simeq 50 \lambda_X^4 ({\rm 1TeV}/m_{\tilde{X}})^4 (m/{\rm 1TeV})^2$ (pb), so that the symmetric part of the dark matter is annihilated away naturally. the Surviving one has a role of the dark matter observed today. Using the asymptotic value of the yield of the asymmetry $[a_{DM}/s]_\infty$, the abundance of the dark matter in the present universe is given by
\begin{eqnarray}
\Omega_{DM} h^2
\simeq
\frac{m [a_{DM}/s]_\infty s_0}{\rho_c/h^2},
\end{eqnarray}
where $s_0 \simeq 2890$ is the entropy density of the present Universe, and $\rho_c/h^2 \simeq 1.05 \times 10^{-5}$ (GeV/cm$^{-3}$) is the critical density. On the other hand, the baryonic abundance is determined by the $B-L$ asymmetry. When the temperature of the Universe becomes as low as 100 GeV, the sphaleron process is frozen out, and, after that, $B$ and $L$ are conserved individually. The relation between $B$ and $B-L$ after taking account of finite mass effects is given by $B/(B-L)\simeq 0.31$~\cite{Kaplan:2009ag}. Using the asymptotic value of the yield $[a_{BL}/s]_\infty$, the baryonic abundance in the present universe is determined by
\begin{eqnarray}
\Omega_{B} h^2
\simeq
\frac{0.31 m_p [a_{BL}/s]_\infty s_0}{\rho_c/h^2},
\end{eqnarray}
where $m_p \simeq$ 938 MeV is the mass of a nucleon (proton).

\begin{figure}[t]
\begin{center}
\includegraphics[scale=0.8]{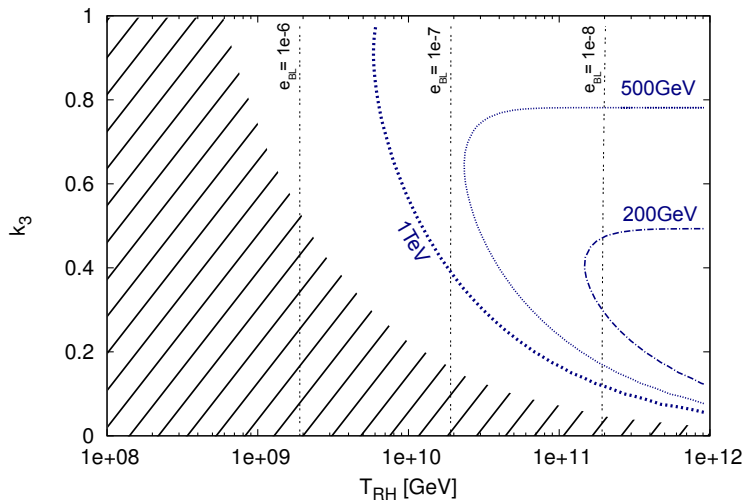}
\caption{\small Dark matter mass as a function of $T_{\rm RH}$ and $\kappa_3$. At each point, black dashed lines show $\epsilon_{BL}$ giving the correct baryon asymmetry $a_B/s \simeq 8.3 \times 10^{-11}$, while blue lines show the dark matter mass $m$ giving the correct dark matter abundance $\Omega_{DM}h^2 \simeq 0.110$.
In the shaded region, the relic abundance of the dark matter cannot be explained by asymmetric dark matter scenario, because the dark matter annihilation cross section is suppressed due to large dark matter mass.
}
\label{DM mass}
\end{center}
\end{figure}

In Fig.\ref{DM mass}, we show the contour plot of the dark matter mass $m$ which gives a correct value of $\Omega_{DM}h^2$ observed by the WMAP experiment ($\Omega_{DM}^{({\rm exp.})} \simeq 0.110/h^2$). The plot is shown on the $(T_{\rm RH}, \kappa_3)$-plane, where the reheating temperature is defined by $T_{\rm RH} = 0.1 (\Gamma M_{\rm pl})^{1/2}$. On the other hand, $\epsilon_{BL}$ is set to give the correct baryonic abundance ($\Omega_B^{({\rm exp.})} \simeq 0.0227/h^2$). Other parameters are fixed to be the same as those in the previous figure. It can be seen that the mass of dark matter $m$ can be of the order of the TeV scale. This fact is very welcome because the supersymmetric mass term of the dark matter field $mX\bar{X}$ may have the same origin as that of the $\mu$ term (supersymmetric Higgsino mass term). In the plot, the value of $\epsilon_{BL}$ is also shown. The definition of the parameter in Eq.(\ref{ASM}) suggests that $\epsilon_{BL}$ takes a value of $\sim 10^{-6}$ when $\kappa \sim 1$ and $y \sim 0.1$ with the complex phase of $y_i$ being ${\cal O}(0.1-1)$. Interestingly, it can be found that the parameter of this order is consistent with the dark matter with the TeV-scale mass.

\subsection{Direct dark matter production from $\tilde N_1$ decay}

\begin{figure}[t]
\begin{center}
\includegraphics[scale=0.8]{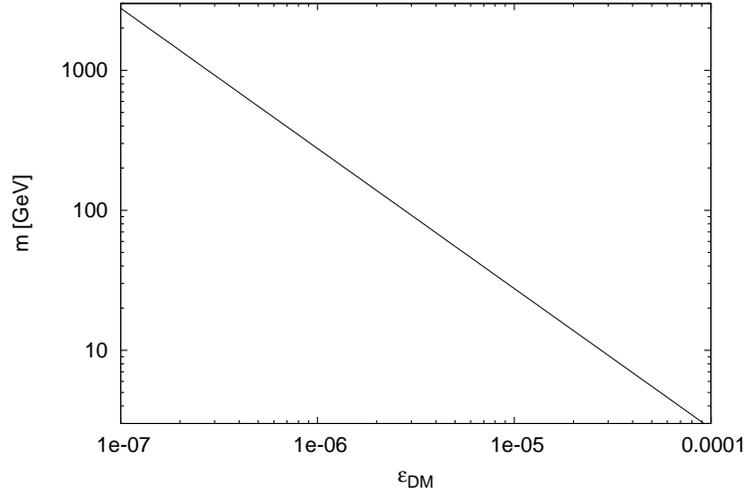}
\caption{\small The mass of the dark matter giving the correct value of the dark matter abundance with $T_{\rm RH} = 10^7$ GeV. We set $\epsilon_{BL} = 1.9\times 10^{-4}$, which gives $a_B/s\simeq 8.3\times 10^{-11}$.}
\label{nonzeroedm}
\end{center}
\end{figure}

In the previous subsection, we have assumed that the dark matter asymmetry is originally coming from the inflaton decay into leptons ($\epsilon_{BL} \gg \epsilon_{DM}$) through the ADM messenger interaction. However, it is also possible to consider the case where the asymmetry comes directly from the inflaton decay ($\epsilon_{BL} \sim \epsilon_{DM})$. This possibility is also of interest, because the dark matter asymmetry can be as large as the $B-L$ asymmetry even if $T_{\rm RH}$ is low. The washout effect and the ADM messenger interaction are not active in such a low reheating temperature because the reaction rates of these processes are proportional to $T_{\rm RH}^3$, while that of the inflaton decay is proportional to $T_{\rm RH}$. As a result, the produced asymmetries $a_{BL}/s$ and $a_{DM}/s$ are determined by $\epsilon_{BL}$ and $\epsilon_{DM}$, respectively. This is also a simultaneous generation of baryon and dark matter asymmetries, which is different from the conventional ADM~\cite{Kaplan:2009ag} and the darkogenesis (baryogenesis from dark sector)~\cite{Shelton:2010ta, Haba:2010bm} scenarios. In Fig.\ref{nonzeroedm}, the dark matter mass $m$ giving the correct value of $\Omega_{DM}h^2$ with $T_{RH}=10^7$~GeV is shown as a function of $\epsilon_{DM}$, where $\epsilon_{BL}$ is set to be $1.9\times 10^{-4}$ which gives $a_B/s \sim 8.3\times 10^{-11}$. It can be seen that a suitable number of $a_{DM}/s$ can be produced with the TeV-scale dark matter even if the reheating temperature is small.

\section{Conclusions and discussions}

We have proposed a simple mechanism for generating the suitable baryon asymmetry and dark matter energy density through the decay of inflaton, which is identified as the right-handed sneutrino.
At first, the lepton asymmetry is generated by the decay of the inflaton and then the asymmetry is transferred into the suitable dark matter asymmetry through the ADM interaction.
We have simply added the dark and antidark matter fields to the Lagrangian of the sneutrino inflation and then the most important ingredient of the ADM scenario, namely the ADM messenger interaction can be derived naturally as a nonrenormalizable operator suppressed by an appropriate energy scale.
As a result, important cosmological issues (inflation, dark matter and baryon mass densities of the universe) and neutrino masses and mixing have the same origin in the model of sneutrino inflation with the ADM, that is, the {\it unification} of inflation, baryogenesis, and dark matter.
Since the proposed model is written in terms of a renormalizable theory, it is regarded as a UV-completion of the ADM scenario.

Finally, we comment on neutrino oscillation, which requires suitable
 neutrino mass hierarchy and mixings among three generations.
As for mass hierarchy,
 neutrino Yukawa couplings in our model are hierarchical as
 ${\rm Max.} [h^{(\nu)}_{1i}]\sim 10^{-3}$ and
 ${\rm Max.} [h^{(\nu)}_{ij}] \sim 0.1$ ($i,j = 2,3$),
 where index ``1" denotes the lightest neutrino field.
Therefore ``1" can be the 1st or 3rd generation, which
 corresponds to normal hierarchy or inverted hierarchy
 neutrino mass eigenvalues, respectively.
(Notice that a degenerate neutrino is not allowed.)
As for generation mixing,
 we can obtain a suitable MNS matrix by taking into
 account mixings from charged lepton side.
(Un)fortunately, there are undetermined degrees of freedom
 in Yukawa couplings of charged lepton in our setup.

\vspace{1cm}

{\large \bf Acknowledgments}\\

\noindent
This work is partially supported by Scientific Grant by Ministry of Education and Science, Nos. 20540272, 20039006, 20025004, 21740174, and 22244021. We also thank the Yukawa Institute for Theoretical Physics at Kyoto University (YITP). Discussions during the YITP workshop ``Summer Institute 2010'' (YITP-W-10-07) were useful to develop this work. This work was supported by the World Premier International Research Center Initiative (WPI Initiative), MEXT, Japan. The work of RS is supported in part by JSPS Research Fellowships for Young Scientists.

\end{document}